\documentclass[12pt,preprint]{emulateapj}

\usepackage{amsmath}

\slugcomment{}

\shorttitle{Chemical composition of Titan's lakes}

\shortauthors{Cordier et al.}

\begin{document}

\title{An estimate of the chemical composition of Titan's lakes}

\author{
Daniel~Cordier,\altaffilmark{1,2,3}
Olivier~Mousis,\altaffilmark{4,5}
Jonathan I.~Lunine,\altaffilmark{4}
Panayotis~Lavvas,\altaffilmark{4}
and V{\'e}ronique Vuitton\altaffilmark{6}}

\email{daniel.cordier@ensc-rennes.fr}

\altaffiltext{1}{Ecole Nationale Sup{\'e}rieure de Chimie de Rennes, CNRS, UMR 6226, Avenue du G\'en\' eral Leclerc, CS 50837, 35708 Rennes Cedex 7, France}
                 
\altaffiltext{2}{Universit\'e europ\'{e}enne de Bretagne}
                 
\altaffiltext{3}{Institut de Physique de Rennes, CNRS, UMR 6251, Universit{\'e} de Rennes 1, Campus de Beaulieu, 35042 Rennes, France}

\altaffiltext{4}{Lunar and Planetary Laboratory, University of Arizona, Tucson, AZ, USA}

\altaffiltext{5}{Universit{\'e} de Franche-Comt{\'e}, Institut UTINAM, CNRS/INSU, UMR 6213, 25030 Besan\c{c}on Cedex, France}

\altaffiltext{6}{Universit{\'e} Joseph Fourier, Laboratoire de Plan{\'e}tologie de Grenoble, CNRS/INSU, France}

\begin{abstract}
{Hundreds of radar-dark patches interpreted as lakes have been discovered in the north and south polar regions of Titan. We have estimated the composition of these lakes by using the direct abundance measurements from the Gas Chromatograph Mass Spectrometer (GCMS) aboard the Huygens probe and recent photochemical models based on the vertical temperature profile derived by the Huygens Atmospheric Structure Instrument (HASI). Thermodynamic equilibrium is assumed between the atmosphere and the lakes, which are also considered as nonideal solutions. We find that the main constituents of the lakes are ethane (C$_2$H$_6$) ($\sim$ 76--79\%), propane (C$_3$H$_8$) ($\sim$ 7--8\%), methane (CH$_4$) ($\sim$ 5--10\%), hydrogen cyanide (HCN) ($\sim$ 2--3\%), butene (C$_4$H$_8$) ($\sim$ 1\%), butane (C$_4$H$_{10}$) ($\sim$ 1\%) and acetylene (C$_2$H$_2$) ($\sim$ 1\%). The calculated composition of lakes is then substantially different from what has been expected from models elaborated prior to the exploration of Titan by the Cassini-Huygens spacecraft.}
\end{abstract}


\section{Introduction}
\label{sec:intro}

The surface of Saturn's haze-shrouded moon Titan had long been proposed to have oceans or seas, on the basis of the stability of liquid methane and ethane at the ground level (Flasar 1983; Lunine et al. 1983; Lorenz et al. 2003). Ground-based radar observations  ruled out the presence of a global ocean in the 1990s (Muhleman et al. 1995), but the presence of isolated lakes was not precluded (Campbell et al. 2003). A large, dark, lake-like feature subsequently named Ontario Lacus was detected at Titan's south polar region by the Cassini ISS system in 2005 (McEwen et al. 2005) and hundreds of radar dark features with a variety of properties consistent with liquid-filled lakes were found in the northern hemisphere by the Cassini RADAR system (Stofan et al. 2007). 

The chemical composition of the lakes of Titan is still not well determined. Good quality spectral data of the Ontario Lacus have been obtained by the Visual and Infrared Mapping Spectrometer (VIMS) aboard Cassini but the only species that seems firmly identified is C$_2$H$_6$ (Brown et al. 2008); the atmosphere contains so much CH$_4$ that it is very difficult to detect the surface liquid phase of this molecule even if it is dominant in the lakes. Because the detection of other compounds in the lakes of Titan remains challenging in the absence of in situ measurements, the only way to get a good estimate of the chemical composition of these lakes is to elaborate a thermodynamic model based on theoretical calculations and laboratory data. Several models, that investigate the influence of photochemistry and the atmospheric composition on the chemical composition of liquids formed on the surface of Titan, have been elaborated in the pre-Cassini years (Lunine et al. 1983; Dubouloz et al. 1989; McKay et al. 1993; Tokano 2005). Based on atmospheric observations these models assumed surface bodies of liquid on Titan to contain a mixture of C$_2$H$_6$, CH$_4$ and N$_2$ and a large number of dissolved minor species. 

\begin{table}[h]
\label{atmoscompo}
\caption[]{Assumed composition of Titan's atmosphere at the ground level.}
\begin{center}
\begin{tabular}{lcc}
\hline
\hline
\noalign{\smallskip}
Atmosphere			& Mole fraction			    		& Determination         \\
\hline
H$_2$				& $9.8 \times 10^{-4}$			& Huygens GCMS$^{(a)}$	\\
N$_2$				& 0.95                          			& This work				\\
CH$_4$				& 0.0492 						& Huygens GCMS$^{(b)}$	\\
CO					& $4.70 \times 10^{-5}$			& Cassini CIRS$^{(c)}$  \\
$^{40}$Ar				& $4.32 \times 10^{-5}$ 	    		& Huygens GCMS$^{(b)}$	\\
C$_2$H$_6$          		& $1.49 \times 10^{-5}$         		& This work				\\
\hline
\end{tabular}
\tablecomments{$^{(a)}$Owen \& Niemann 2009; $^{(b)}$Niemann et al. 2005; $^{(c)}$De Kok et al. 2007. N$_2$ and C$_2$H$_6$ abundances are determined from our model (see text).}
\end{center}
\end{table}

However,  Cassini-Huygens measurements have improved our knowledge of the structure and composition of Titan's atmosphere, requiring the solubilities to be recomputed under actual Titan conditions. In particular, the Gas Chromatograph Mass Spectrometer (GCMS) aboard Huygens and the Cassini Composite Infrared Spectrometer (CIRS) provided new atmospheric mole fraction data (see Table 1 and Niemann et al. 2005). Moreover, near-surface brightness temperatures at the high latitudes where the lakes exist have now been determined (Jennings et al. 2009). 

Here, we propose a model that takes into account these recent advances and thus provides the most up-to-date chemical composition of Titan's lakes as a function of their location on the satellite's surface. Our model considers the same assumptions as those made by Dubouloz et al. (1989) (hereafter D89) when they calculated the composition of the hypothetical ocean proposed to exist on Titan in the years prior the Cassini-Huygens exploration. The lakes are then considered as nonideal solutions in thermodynamic equilibrium with the atmosphere. This assumption is supported by recent calculations who showed that raindrops could reach the ground in compositional equilibrium with the atmosphere (Graves et al. 2008).

\section{The model of lake-atmosphere equilibrium}
\label{sec:model}

 The general phenomenological picture of our model is the following: high altitude photochemistry produces 
gases and aerosols in Titan's atmosphere, which are transported to lower altitude via atmospheric mixing, molecular
diffusion, and sedimentation. At low temperature conditions found in the lower stratophere and troposphere,
the formed gases can condense on the aerosols and their precipitation rates (see Table 2).
are given by photochemical models (Lavvas et al. 2008a, 2008b; Vuitton et al. 2008). From the different vapor pressures of the molecules that condense in Titan's atmosphere, these models
define the altitude (\textit{i.e.} temperature) at which the condensation of each gas will start.
Although the condensed phase can start as a liquid, the temperature of the surface is smaller than the temperature at the altitude 
where condensation begins for some of the molecules. This allows some of the condensables to reach the surface in the solid phase. 
The physical state of these precipitaties as they enter the seas at Titan's surface level does not matter as they are
assumed to dissolve in thermodynamic equilibrium.
The considered gas phase is representative of Titan's atmosphere (see Table 1) and the ethane mole fractions in liquid and gas are considered as unknowns of the problem. Mole fractions of other species present in precipitation are supposed to have negligible gas phase abundances. We place our model in the framework of the regular solution theory. Thus, because the thermodynamic equilibrium is assumed between lakes and atmosphere, the equality of chemical potentials for each species listed in Table 1 can be written as (Eq. 1 of D89):

\begin{equation}\label{equa1}
  Y_{i} \, P = \Gamma_{i} \, X_{i} \, P_{vp,i},
\end{equation}

  where $P$ is the total pressure at Titan's surface, $Y_{i}$ and $X_{i}$ are respectively the mole fractions of the $i$ compound in the atmosphere and the liquid, $P_{vp,i}$ its vapor pressure, and $\Gamma_{i}$ its activity coefficient in the liquid given by Eq. 2 of D89. Abundances of compounds below C$_2$H$_6$ in Table 2 are expressed proportionally to that of C$_2$H$_6$ both in the precipitation and in the lakes. Because the system of involved equations is non-linear, it is solved with the use of the Newton Raphson's method.

Our model also allows us to estimate the fractions of solid precipitates that can be dissolved in the lakes of Titan. To this end, we calculate
the \textit{saturation} mole fraction\footnote{The saturation mole fraction of the $i$ compound corresponds to the maximum mole fraction of $i$ in the liquid form. Above this value, the $i$ material in excess remains in solid form.}
$X_{i,sat}$ of the $i$ compound, which is given by (Eq. 7 of D89):

\begin{equation}\label{equa2}
 \mathrm{ln}(\Gamma_{i} \,  X_{i,sat}) = (\Delta H_{m}/R T_{m})(1-T_{m}/T),
\end{equation}

\noindent where $T_m$ is the component's melting temperature and $\Delta H_m$ its enthalpy of fusion. Our calculation procedure is then conducted as follows: 

\begin{enumerate}
    \item The unknown $X_{i}$'s and $Y_{i}$'s are computed via the Newton-Raphson method.
    \item Once the $X_{i}$'s have been determined, the $X_{i,sat}$'s are in turn calculated and compared to the $X_{i}$'s for each species. 
           If for compound $i$ we get $X_{i,sat}~<~X_{i}$, then we fix $X_{i} = X_{i,sat}$.
    \item We get new values of $X_{i}$'s and $X_{i,sat}$'s via the resolution of the nonlinear system.
    \item The iterations are continued until we get a difference between $X_{i,sat}$ and $X_{i}$ lower than $10^{-6}$,
          value for which the numerical inaccuracy is clearly negligible compared to other sources of uncertainties.
\end{enumerate}

\noindent The known $Y_{i}$'s are given in Table~\ref{atmoscompo}. The precipitation rates used here are given in Table~\ref{prodrate} and derive from the photochemical models of Lavvas et al. (2008a,b) and Vuitton et al. (2008) and correspond to the main products of CH$_4$ and N$_2$ photolysis. These rates allow to express each $i$ compound that precipitates in the form $X_{i}$~=~$\frac{\tau_i}{\tau_{C_2H_6}}~\times$~$X_{C_2H_6}$. 
We also ensure that $\sum_{i} X_i$ = 1 and $\sum_{i} Y_i$ = 1. The thermodynamic data used in our calculations derive from the NIST database\footnote{\texttt{http://webbook.nist.gov}} when they are available and the remaining ones have been taken from D89. Note that H$_2$ is the only compound whose mole fraction in the liquid is not determined with the aforementioned procedure. Instead, we calculate the amount of dissolved H$_2$ in the liquid via Henry's law (D89).

\begin{table}[h]
\caption[]{Precipitation rates ($\tau$) assumed at the ground level.}
\begin{center}
\begin{tabular}{lcc}
\hline
\hline
\noalign{\smallskip}
				& Compound		& Rate $\tau$ (cm$^{-2}$ s$^{-1}$) 	\\
\hline
Liquids			& C$_2$H$_6$	& $3.4 \times 10^{9}$$^{(a)}$	\\
				&C$_3$H$_8$		& $3.3 \times 10^{8}$$^{(a)}$ 	\\	
				&C$_4$H$_8$          & $6.2 \times 10^{7}$$^{(a)}$	\\
\noalign{\smallskip}
\hline
Solids 			&HCN			& $1.3 \times 10^{8}$$^{(a)}$	\\
				&C$_4$H$_{10}$	& $5.4 \times 10^{7}$$^{(a)}$	\\
				&C$_2$H$_2$		& $5.1 \times 10^{7}$$^{(a)}$	\\
				&CH$_3$CN		& $4.4 \times 10^{6}$$^{(a)}$	\\
				& CO$_2$		& $1.3 \times 10^{6}$$^{(a)}$	\\
				& C$_6$H$_6$	& $1.0 \times 10^{6}$$^{(b)}$	\\

\hline
\end{tabular}
\tablecomments{$^{(a)}$Lavvas et al. (2008a, 2008b); $^{(b)}$Vuitton et al. (2008).}
\end{center}
\label{prodrate}
\end{table}

\section{Results}
\label{sec:results}

  Our calculations in the framework of thermodynamic equilibrium have been performed for two different regions of Titan's surface. The first zone corresponds to the vicinity of the landing site of the Huygens probe, where the surface temperature was measured to be $93.65$ K (Niemann et al. 2005). The Huygens probe detected drainage-like features and a high surface relative humidity, so the presence of liquids cannot be excluded in this area (Tomasko et al. 2005; Niemann et al. 2005). The second zone corresponds to the north pole of Titan where the surface temperature is around $\sim$ 90 K based on near-surface brightness temperature measurements (Jennings et al. 2009). In both cases, the atmospheric pressure is assumed to be identical and corresponds to that (1.46 bar) measured by Huygens at the ground level (Niemann et al. 2005).

\begin{figure}
\begin{center}
\includegraphics[angle=0,width=8.5cm]{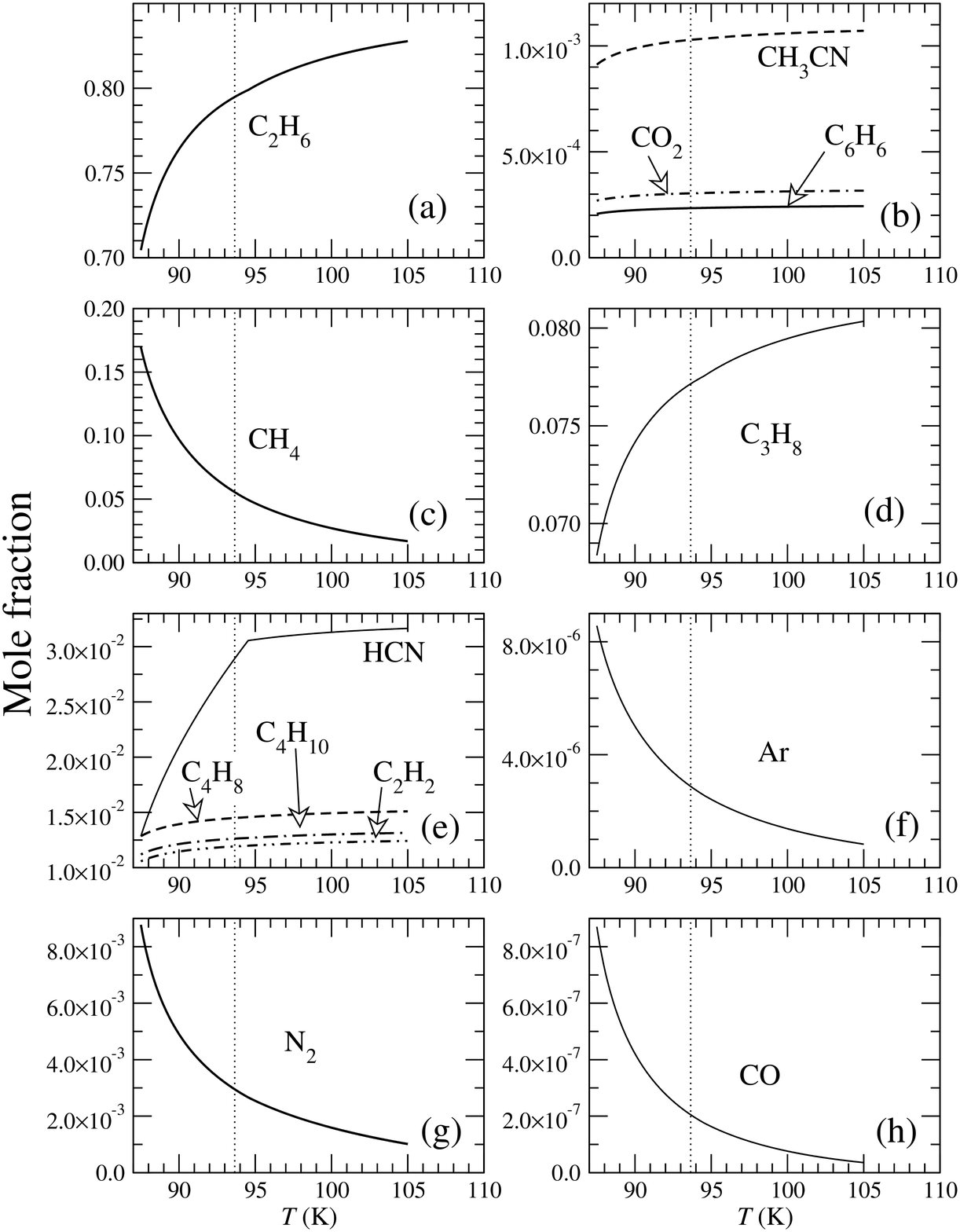}
\caption{\label{influ_T}(a)--(h): composition of lakes as a function of the surface temperature. The vertical dashed line corresponds to the surface temperature of $93.65$ K measured by Huygens.}
\end{center}
\end{figure}

\begin{figure}
\begin{center}
\includegraphics[angle=0,width=8.5cm]{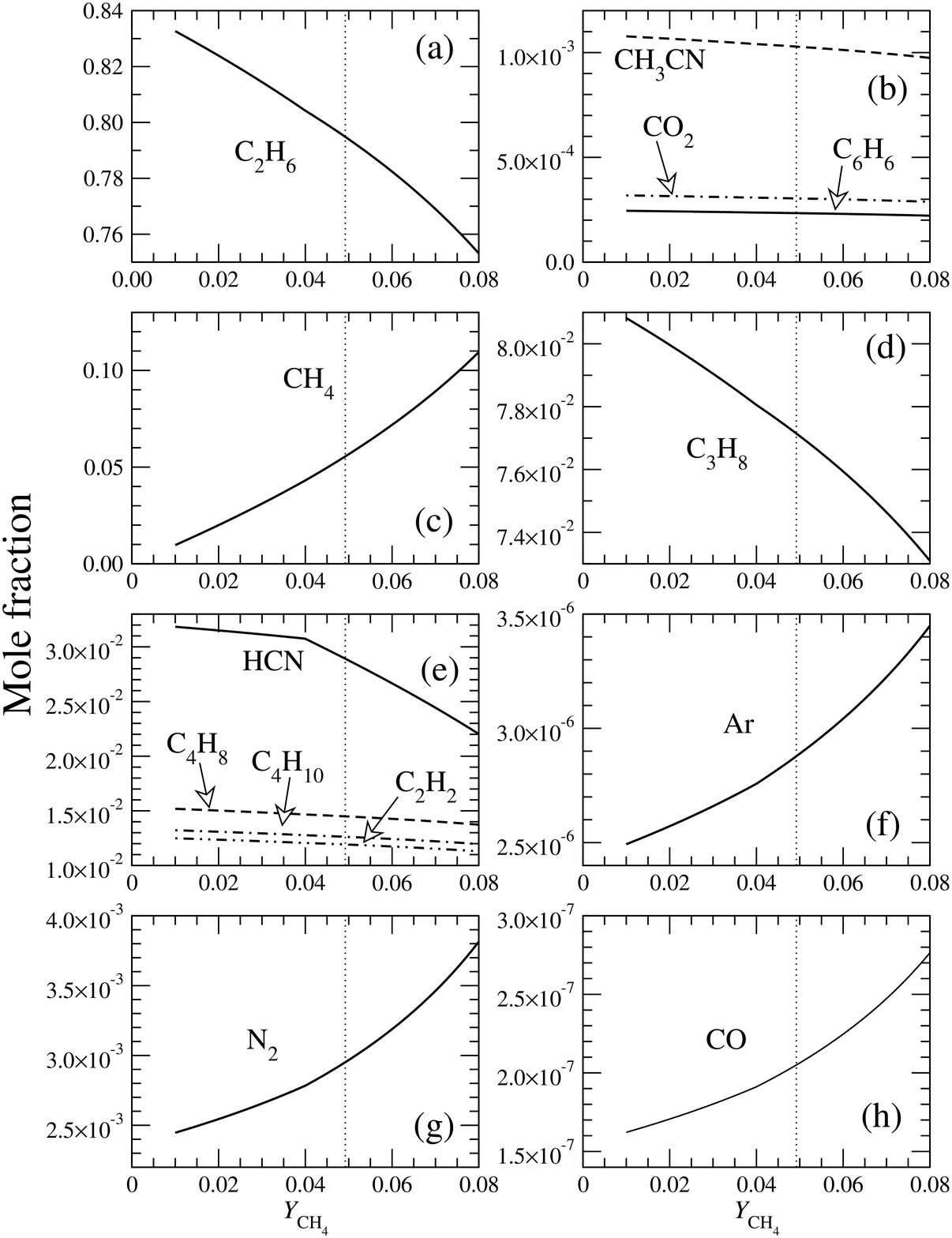}
\caption{\label{influ_CH4}(a)--(h): composition of lakes as a function of the methane atmospheric mole fraction, assuming a surface temperature of $93.65$ K. The vertical dashed line corresponds to the methane atmospheric mole fraction measured by Huygens at the ground level.}
\end{center}
\end{figure}

Figure~\ref{influ_T} shows the variation of the composition of Titan's lakes as a function of the surface temperature. It appears that the mole fractions of CH$_4$, N$_2$, CO and Ar decrease with the increase of temperature, while the mole fraction of C$_2$H$_6$ and of the precipitates increase. This is due to the vapor pressure of C$_2$H$_6$, whose temperature dependence is lower than those of CH$_4$, N$_2$, CO and Ar. Figure~\ref{influ_CH4} displays the variation of the composition of the lakes of Titan as a function of the atmospheric CH$_4$ mole fraction at the ground level, assuming a surface temperature of $93.65$ K. It shows that an increase of the CH$_4$ atmospheric mixing ratio enhances its corresponding mole fraction in the liquid, as well as those of N$_2$, CO and Ar. Interestingly enough, HCN is the only compound reaching saturation within the considered ranges of temperature and atmospheric methane mole fraction. Indeed, the HCN mole fraction represented as  a function of the soil temperature in figure~\ref{influ_T} panel (e) corresponds to the saturation limit for any temperature lower than 94.5 K (change of the curve's slope). This is due to the solubility of this compound diminishing with decreasing temperature. Moreover, 
figure~\ref{influ_CH4} panel (c) shows that when $Y_{\rm CH_{4}}$ is larger than $\sim 0.04$, the lake is saturated in HCN. Because it is more dense than the liquid\footnote{The mean molar volume of the liquid HCN is $4.8 \times 10^{-5}$ m$^{3}/$mol while that of solid HCN is lower than $3.8 \times 10^{-5}$ m$^{3}/$mol.}, the non-dissolved HCN should sink in the lakes of Titan.

Table \ref{nomcompo} gives the mole fractions of the main compounds in lakes formed on the surface of Titan and shows that, 
whatever the considered site, their composition is dominated by C$_2$H$_6$, C$_3$H$_8$, CH$_4$, HCN, C$_4$H$_8$, C$_4$H$_{10}$ and C$_2$H$_2$. On the other hand, with mole fractions much lower than 1\%, N$_2$, C$_6$H$_6$, CH$_3$CN, CO$_2$, Ar, CO and H$_2$ are found to be minor compounds in the lakes.

\section{Discussion}
\label{sec:discussion}

The use of more up-to-date thermodynamic data and the recent Cassini-Huygens measurements, results in our calculated composition of the lakes differing from that of the hypothetical global ocean determined by D89. Indeed, considering the D89 case most closely corresponding to actual Titan conditions ($T= 92.5$ K, $Y_{\rm Ar}=0$ and $Y_{\rm CH_{4}}= 0.0492$; see their Fig.~1), they obtained mole fractions of C$_2$H$_6$, CH$_4$ and N$_2$ of $\sim$ 0.35, 0.60 and 0.05, respectively. The most striking difference in the comparison is that the mole fraction of N$_2$ is more than 10 times lower in our results than in theirs, and some compounds considered as minor in the ocean of D89 are not negligible in the composition of our lakes. However, our model reproduces the results of D89 when we adopt precisely the same thermodynamic parameters. The fact that relatively small differences (order of a few percent) between our thermodynamic data extracted from the NIST database and those of D89 lead to significantly different results illustrates the non-linearity of the system of equations determining the composition of the liquid.  It is difficult to quantify the errors on the predicted mole fractions of the different species because our model is based on thermodynamic data much of which is provided either without uncertainties or is derived from extrapolation of data to low temperature.
   However, numerical tests aiming to investigate the sensitivity of our model 
to the photochemical models we employ indicate that a $\pm$ 30\% variation of the C$_2$H$_6$ precipitation rate induces similar variations for the C$_3$H$_8$ to C$_6$H$_6$ lake mole fractions but poorly alters the N$_2$ to C$_2$H$_6$ lake mole fractions given in Table 3. Moreover, as shown by Fig. 2, a variation of the methane atmospheric abundance hardly affects the lake mole fractions of the precipitates. Pure numerical errors due to the algorithm have been estimated absolutely negligible.

We also outline that the assumption of thermodynamic equilibrium between the lakes and the atmosphere is a crude approximation since in fact the humidity and temperature of the atmosphere in contact with the lakes is determined by dynamic processes such as convection and wind advection (Mitri et al.  2007). 

Our solubility calculations imply that a number of species produced by methane photolysis and energetic particle chemistry in Titan's upper atmosphere should be readily detectable with a mass spectrometer carried to the surface of a liquid-filled lake by a Huygens-like entry probe (Coustenis et al. 2009). The measured abundances of multiple minor constituents in the lake, coupled to measurements and models of stratospheric abundances and production rates, and direct temperature measurements of the lake surface, will constrain lake properties that are of interest in understanding the methane hydrologic cycle. For example, at the winter pole a seasonally deposited upper-layer of liquid methane might exist on top of a longer-lived ethane-methane liquid reservoir by virtue of methane's lower density and limited vertical mixing in the cold lakes (Stevenson and Potter 1986). Such a transient layer would be bereft of minor components compared with our values thanks to the slow sedimentation rate of the high altitude aerosols compared to the seasonal (meteorological) methane deposition rate; our solubility values provide a means of calculating the extent to which the longer-lived liquid reservoir below has mixed into the methane meteorological layer.  (The extreme cold of the tropopause of Titan prevents the hydrocarbon constituents other than methane and possibly ethane from passing directly to the lower atmosphere in the gas phase; thus the lakes must be seeded by stratospheric aerosol sedimentation).  

\begin{table}[h]
\caption[]{Chemical composition of lakes at the poles and the equator.}
\begin{center}
\begin{tabular}{lcc}
\hline
\hline
\noalign{\smallskip}
 					&  Equator (93.65 K)					& Poles (90 K)			\\
\hline
\multicolumn{3}{l}{Main composition (lake mole fraction)}								\\
N$_2$     				&  $ 2.95\times 10^{-3}$  			& $4.90\times 10^{-3}$ 		\\
CH$_4$   				&  $5.55\times 10^{-2}$  			& $9.69\times 10^{-2}$  		\\
Ar     					&  $2.88\times 10^{-6}$ 			& $5.01\times 10^{-6}$ 		\\
CO     				    &  $2.05\times 10^{-7}$  			& $4.21\times 10^{-7}$  		\\
C$_2$H$_6$   			&  $7.95\times 10^{-1}$  			& $7.64\times 10^{-1}$  		\\
C$_3$H$_8$   			&  $7.71\times 10^{-2}$  			& $7.42\times 10^{-2} $  		\\
C$_4$H$_8$   			&  $1.45\times 10^{-2}$  			& $1.39\times 10^{-2}$  		\\
H$_2$				    &  $5.09\times 10^{-11}$			& $3.99\times 10^{-11}$		\\
\hline 
\multicolumn{3}{l}{ Solutes (lake mole fraction)}								\\
HCN    				& $2.89\times 10^{-2}$	(s)   		& $2.09\times 10^{-2}$	(s) 	\\
C$_4$H$_{10}$ 		& $1.26\times 10^{-2}$ (ns)  		& $1.21\times 10^{-2}$	(ns)	\\
C$_2$H$_2$   		& $1.19\times 10^{-2}$ (ns) 		& $1.15\times 10^{-2}$	(ns) 	\\
C$_6$H$_6$   		& $2.34\times 10^{-4}$ (ns)  	    & $2.25\times 10^{-4}$	(ns)    \\
CH$_3$CN  			& $1.03\times 10^{-3}$ (ns)  		& $9.89\times 10^{-4}$	(ns) 	\\
CO$_2$    			& $3.04\times 10^{-4}$ (ns)  		& $2.92\times 10^{-4}$	(ns) 	\\
\hline
\end{tabular}
\tablecomments{(s): saturated; (ns) non saturated.}
\end{center}
\label{nomcompo}
\end{table}

A longer-lived ethane-methane lake sampled at the summer pole might still have undergone pole-to-pole transport on timescales of tens of millenia thanks to the precession of perihelion of Saturn's orbit around the Sun (Aharonson et al. 2009).
The abundances of minor constituents compared to our values which assume accumulation over geologic time, coupled with the aerosol sedimentation rate, could  be used to ``date'' the liquid reservoir and hence test whether they have been cycled on the ``Milankovitch" timescale. For species that are chemically inactive and occur as gas only in the atmosphere, such as the noble gases, the lakes provide a second reservoir other than the atmosphere to measure abundances. Variations from the atmospheric value of, for example, $^{40}$Ar/$^{36}$Ar, might hint at contact between the lakes and a much deeper crustal reservoir of liquid methane and ethane not in contact with the atmosphere (Hayes et al. 2008).  Finally, our results provide the chemical data needed to compute the amount of deposition of various hydrocarbons and nitriles in fluvial valleys in the Titan's midlatitudes, as a function of the flow of methane runoff from convective storms, allowing potential tests of models of fluvial erosion (Perron et al. 2006). 

\acknowledgements
This work was supported in part by the CNES. Support from the PID program ``Origines des Plan{\`e}tes et de la Vie" of the CNRS, and the {\it Cassini} project, are also gratefully acknowledged. We acknowledge Bruno B{\'e}zard, S{\'e}bastien Lebonnois and Pascal Rannou for helpful discussions about Titan's atmosphere. We thank John Prausnitz for providing us with some unpublished thermodynamic data.
We thank an anonymous reviewer for his constructive comments which helped us improve our manuscript.


\end{document}